# Broadband continuous spectral ghost imaging for high resolution spectroscopy

Caroline Amiot[1], Piotr Ryczkowski[1], Ari T. Friberg[2], John M. Dudley[3], and Goëry Genty[1*]

[1]Laboratory of Photonics, Tampere University of Technology, Tampere, Finland

[2]Department of Physics and Mathematics, University of Eastern Finland, Joensuu, Finland

[3]Institut FEMTO-ST, UMR 6174 CNRS-Université de Bourgogne-Franche-Comté, Besançon, France

caroline.amiot@tut.fi

piotr.ryczkowski@tut.fi

ari.friberg@uef.fi

john.dudley@univ-fcomte.fr

goery.genty@tut.fi (+358(0)50 346 3069)

*corresponding author

**Ghost imaging is an unconventional imaging technique that generates high resolution images by correlating the intensity of two light beams, neither of which independently contains useful information about the shape of the object [1,2]. Ghost imaging has great potential to provide robust imaging solutions in the presence of severe environmental perturbations, and has been demonstrated both in the spatial [3-10] and temporal domains [11–13]. Here, we exploit recent progress in ultrafast real-time measurement techniques [14] to demonstrate ghost imaging in the frequency domain using a continuous spectrum from an incoherent supercontinuum (SC) light source. We demonstrate the particular application of this ghost imaging technique to broadband spectroscopic measurements of methane absorption, and our results offer novel perspectives for remote sensing in low light conditions, or in spectral regions where sensitive detectors are lacking.**

Ghost imaging is based on correlating two signals: the spatially-resolved intensity pattern of a light source incident on the object (a probe pattern), and the total integrated intensity scattered from the illuminated object. The image is generated by summing over multiple incident patterns, each weighted by the integrated scattered signal. Significantly, this fundamental principle of ghost imaging is not restricted to producing only spatial images, and in a recent time-domain application, measurement of ultrafast signals on picosecond timescales was reported [11–13]. In the frequency domain, the measurement of wavelength-by-wavelength Hanbury Brown-Twiss intensity correlations of a superluminescent laser diode was used to characterize spectral features of chloroform in a proof-of-principle experiment [15]. This approach however used nonlinear detection that requires either a high-power source or very sensitive detectors, and it is also experimentally time consuming as it involves a raster scan approach both in wavelength and time.

In this paper, we introduce a scan-free approach to high-resolution frequency domain ghost imaging that enables the measurement of the spectral transmission (or reflection) of an object using a detector without any spectral resolution. The method has rapid acquisition time, is broadband and requires neither a high-power source nor particularly sensitive detectors. As a particular application, we report spectroscopic measurements of methane absorption lines over a 50 nm bandwidth and with sub-nm resolution. The results are in excellent agreement with independent measurements, and the approach offer great potential for spectral sensing in diffuse light levels and in spectral regions where sensitive spectrometers are not available.

Figure 1 shows the principle of spectral ghost imaging. Consider an object with wavelength-dependent transmission which modulates (or modifies) the spectrum of a broadband light beam interacting with this object. Provided the incident light field exhibits random spectral intensity fluctuations, a ghost image of the object's spectral response can be generated from the normalised correlation function $C(\lambda)$ defined by:

$$C(\lambda) = \frac{\langle \Delta I_{\text{ref}}(\lambda) \cdot \Delta I_{\text{test}} \rangle_N}{\sqrt{\langle \Delta I_{\text{ref}}(\lambda)^2 \rangle \langle \Delta I_{\text{test}}^2 \rangle}}.$$

Physically, $C(\lambda)$ is the wavelength-dependent correlation between multiple measurements of the spectral intensity fluctuations in a reference arm $I_{\text{ref}}(\lambda)$ and the total (or integrated) wavelength-independent intensity $I_{\text{test}}$ in the test arm where light interacts with the object. Here, $\langle \ \rangle_N$ denotes ensemble average over distinct $N$ realisations, and $\Delta I = I - \langle I \rangle_N$.

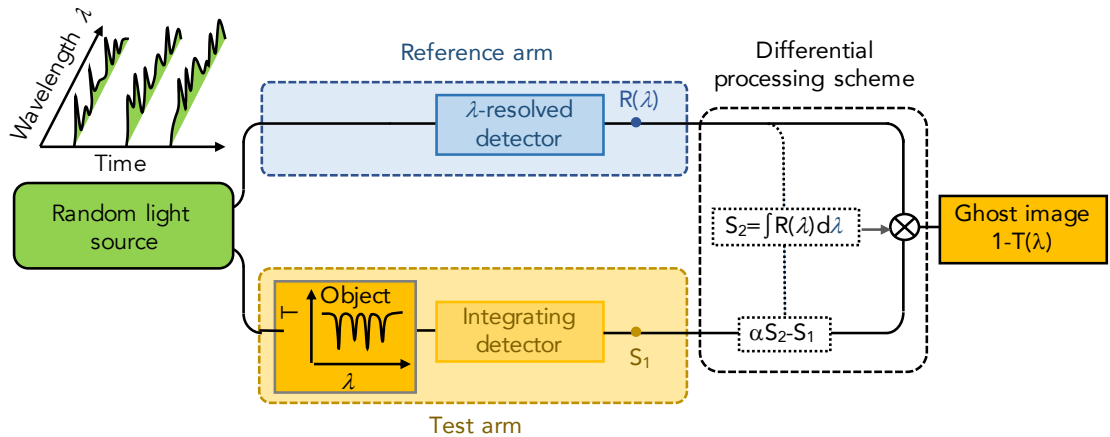

**Fig.1.** Ghost imaging in the spectral domain. The wavelength-dependent transmission T(λ) of an object is obtained from the correlation of the spectrally-resolved fluctuations of a random light source with the spectrally integrated intensity at the object output. Differential detection is numerically implemented to increase the signal-to-noise ratio as shown. Note that the figure considers wavelength-dependent transmission, but the same principle applies for reflection.

Using this approach, we aim to generate a "ghost image" of the absorption spectrum of gas molecules. Figure 2 shows our experimental setup. The light source is a spectrally filtered broadband incoherent supercontinuum [16] extending over 1610–1670nm (see Methods). The supercontinuum source is equally divided between the reference and test arms with a 99/1 fibre coupler. In the reference arm, the spectral fluctuations are measured in real-time using the DFT approach (see Methods) with 0.2 nm resolution, which also sets the resolution of the spectral-domain ghost-imaging scheme. DFT has recently been applied to many studies in ultrafast nonlinear dynamics [17–19], for direct spectroscopic measurements [21–23], label-free molecular identification [24], pump-probe Raman excitation [25], and biomedical diagnostics [26], and is rapidly becoming a standard characterization tool in ultrafast science.

In the test arm, the spectral object is a 16.7 cm long fibre-based single pass methane gas cell (Wavelength References FCS-16-1/4) at atmospheric pressure and room temperature. The absorption lines of the 2v3 overtone transition [27] are in the spectral range of the filtered source, and act as the spectral object modulating the randomly fluctuating incident light. Light transmitted through the cell is measured by an integrating photodetector with 50 ns response time (15 MHz bandwidth, Thorlabs PDA10D2), which, alone, cannot resolve any spectral feature of the gas absorption. Multiple single-shot realisations are recorded by an oscilloscope, and a differential detection scheme is applied in post-processing [28,29] (see Methods).

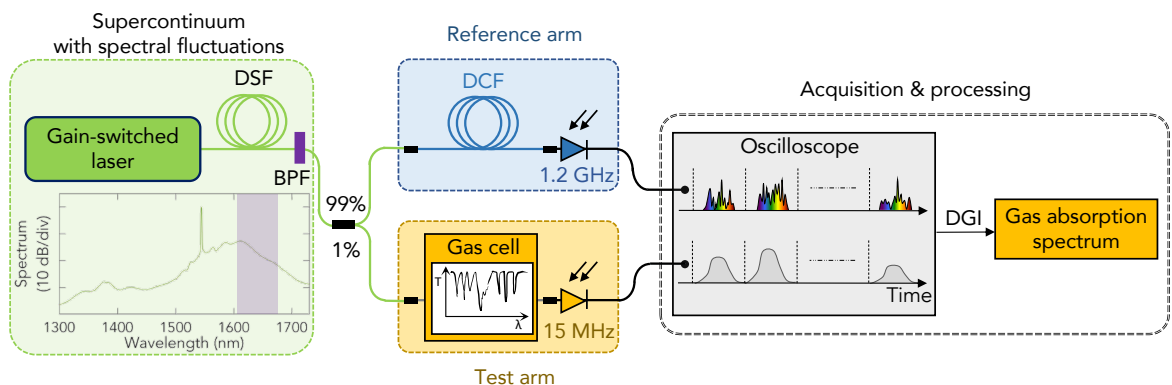

**Fig.2.** Experimental setup. DSF: dispersion-shifted fibre, BPF: band-pass filter, DCF: dispersion compensating fibre. DGI: differential ghost imaging.

Figure 3 shows experimentally recorded spectral fluctuations. In particular, Fig. 3(a) shows four different single-shot spectra measured from the reference arm, clearly illustrating the fluctuations in spectral intensity. The effective bandwidth here is ~0.2 nm as seen in the expanded portion in Fig. 3(b). The average spectral envelope of the filtered spectrum over a large number of consecutive single-shot spectra is also shown in Fig. 3(a), with this envelope being essentially determined by the transmission curve of the 50 nm bandpass filter. We also plot the standard deviation of the fluctuations relative to the average spectrum, which is close to 50% across the full (filtered) bandwidth.

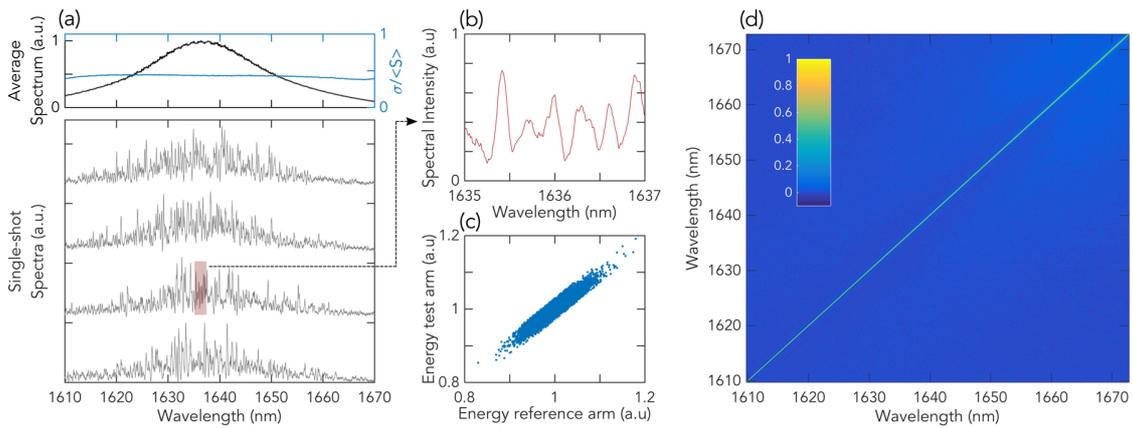

**Fig.3.** Spectral fluctuations of the incoherent supercontinuum source. (a) shows 4 selected examples of recorded single-shot spectra using the DFT technique. The average spectrum and relative standard deviation of the fluctuations are shown in the top panel. (b) is an expanded view of the fluctuations over a 2 nm span shown as the highlighted red rectangle in (a). (c) illustrates the correspondence between the single-shot spectra energy measured from the reference arm and the energy measured at the gas cell output by the integrating detector. (d) Wavelength-correlations of the filtered supercontinuum source measured over 20,000 single-shot spectra.

Ghost imaging relies on the fact that the reference and test beams are perfectly correlated such that a variation measured in the test arm intensity is caused by the change in the fluctuations of the probe in the reference arm. In the case of a sparse object for which the variations in the spectrally integrated transmission are small from one realisation to another, the integrated intensities measured at the output of the test and reference arms should be nearly fully correlated. For our experiments, this is verified explicitly in Fig. 3(c) with a very strong correlation (Pearson correlation coefficient $\rho$ = 0.946). Finally, the random nature of the spectral fluctuations is conveniently visualized in Fig. 3(d) where we plot the wavelength-to-wavelength correlations of a measured ensemble of 20,000 single-shot spectra (see Methods). The plot shows a high degree of correlation only on the diagonal, clearly indicating that there are no spectral correlations within the pulses, a pre-requisite for ghost imaging using random fluctuations.

The ghost spectral image of the methane absorption lines obtained from 20,000 distinct SC spectra is shown in Fig. 4 (solid blue line). For comparison, a direct measurement of the gas absorption using the same filtered supercontinuum source and a conventional optical spectrum analyzer with 0.2 nm resolution is also plotted (solid red line). Both the ghost image and the direct measurement are low-pass filtered to remove the distortion from the supercontinuum spectral envelope (see Methods). We see remarkable agreement over the full 50 nm spectral range of the overtone transitions, with the resolution of the ghost imaging very close to that of the direct measurement. For completeness, we also show in the figure inset the ghost image obtained without the numerical differential scheme implementation and for the same number of 20,000 realisations. No obvious spectral signature of the gas is observed and we can see how the differential post-processing is extremely powerful in revealing the absorption lines with only a limited number of realisations. We emphasize here that the DFT method employed to measure in real time the spectral fluctuations allows for fast data acquisition rates. This means that with the 100 kHz repetition rate SC source used here, it only takes ~200 ms to record the 20,000 realisations used to retrieve the absorption spectrum.

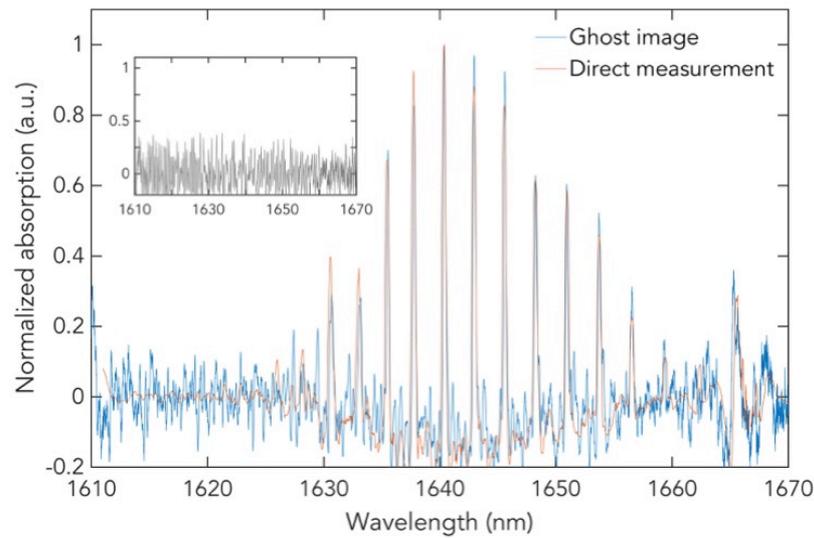

**Fig.4.** Experimental ghost image of the 2v3 overtone transition of methane generated from 20,000 realisations using the numerical differential scheme (solid blue line). A direct measurement performed with the SC source and an optical spectrum anaylzer is shown for comparison (solid red line). The inset plots the experimental ghost image generated from 20,000 realisations without the differential scheme implementation. The ghost image and direct measurement were low-pass filtered to remove the spectral envelope of the SC source.

Significantly, even with a much lower number of realisations, the absorption lines can still be resolved by the ghost measurement. This is illustrated in Fig. 5 where we show the ghost spectral image obtained for an increasing number of realisations. The signal-to-noise ratio increases with the number of measurements, yet one can see how the main absorption lines are already resolved with as little as 2,500 measurements.

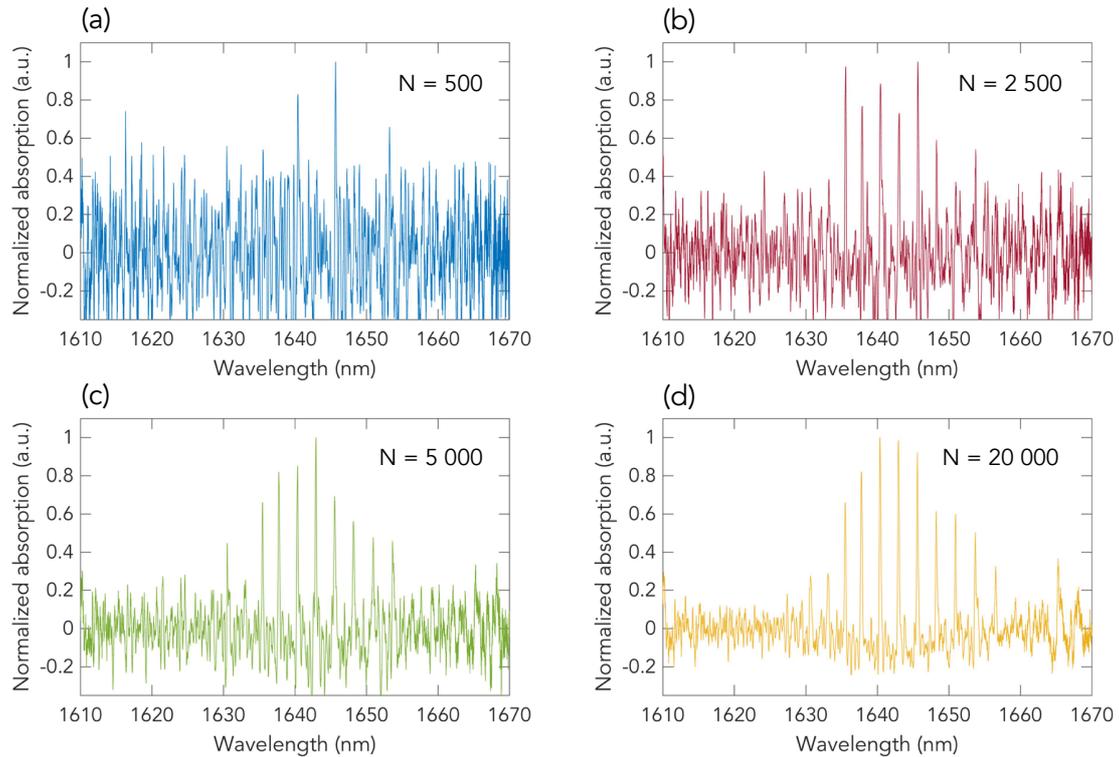

**Fig.5.** Ghost image of the methane absorption lines for an increasing number of realisations as indicated in each sub-panel.

In conclusion, we have demonstrated broadband continuous spectrum ghost imaging using an incoherent supercontinuum light source with large shot-to-shot spectral fluctuations. The method is fast, broadband, scan-free and does not require any high power source or sensitive detector after the object. We have applied the technique to perform high resolution spectroscopic measurements of the absorption spectrum of methane in the 1600 nm wavelength range. The current resolution is limited by the precision with which the linewidth of spectral fluctuations in the reference arm can be measured in real-time, and we anticipate that pm resolution could be reached using a DFT fibre with larger dispersion and/or a tailored supercontinuum source. Finally, it is important to stress that the method is insensitive to linear spectral or temporal distortions occurring between the object and the integrating detector, which would not be the case in a direct measurement configuration [12]. Our results are very significant in showing that spectral ghost imaging has great potential for remote sensing and spectroscopic application in diffuse light conditions or in spectral regions where no sensitive detector exists.

**Methods**

**Incoherent supercontinuum generation** The noisy supercontinuum is generated by launching 1kW, 700ps pulses at 1547nm with 100 kHz repetition rate (Keopsys-PEFL-K09) into a 6-m long DSF with zero-dispersion wavelength at 1510 nm (Corning Inc LEAF). The supercontinuum bandwidth at the output of the DSF extends from ~1300 to over 1700nm. In order to ensure fidelity in the DFT measurements because of attenuation (beyond 1670 nm) and third-order dispersion in the dispersion compensating fibre used for the DFT, the supercontinuum is then filtered in the 1610–1670 nm range with a 50 nm (full-width-at-half-maximum) bandwidth filter (Spectro-on NB-1650-050). The relatively long pump pulses used to generate the supercontinuum leads to significant shot-to-shot spectral fluctuations caused by an initial stage of modulation instability, and it is these random spectral fluctuations that are used in our ghost imaging setup. We also specifically adjust the peak power of the pump pulses injected into the DSF in order to maximize the amplitude of the spectral fluctuations, which reduces the digitization noise on the oscilloscope and yield improved signal-to-noise ratio when measuring in real time these fluctuations with the DFT technique.

**Dispersive Fourier transform** The spectral fluctuations are converted into the time domain by a dispersion compensating fibre (FS.COM customized 150 km dispersion compensating fibre) with total dispersion of 3000psnm−1 and measured with a with a 1.2 GHz InGaAs photodetector (Thorlabs DET01CFC/M) and 20 GHz real-time oscilloscope (Tektronics DSA72004) with 50GS/s sampling rate. The resolution of the DFT is 0.2 nm which sets the resolution of the spectral-domain ghost imaging scheme.

**Differential ghost imaging** The spectral object (the absorption spectrum) is "sparse", i.e. the absorption lines are significantly narrower than the spectral window of the measurement such that the gas molecules only impart very small changes in the total transmitted light. In order to increase the signal-to-noise ratio and reduce the number of measurements, we adopt a differential processing scheme which has the advantage to be sensitive only to the fluctuating part of the transmission. The differential detection is a numerical post-processing scheme originally developed in the spatial domain and which is similar to performing balanced detection in the test arm. It is realized by replacing the integrated intensity in the test arm with a numerically generated differential signal $\alpha \int I_{\text{ref}}(\lambda)d\lambda - I_{\text{test}}$ where: $\alpha = \langle I_{\text{test}} \rangle / \int I_{\text{ref}}(\lambda)d\lambda$.

**Wavelength-to-wavelength correlation of the supercontinuum source** In order to quantify the correlations between any two wavelengths $\lambda_1$ and $\lambda_2$ in the SC, we calculate the spectral correlation matrix given by $\rho(\lambda_1, \lambda_2) = \frac{\langle \Delta I(\lambda_1) \cdot \Delta I(\lambda_2) \rangle_N}{\sqrt{\langle \Delta I(\lambda_1)^2 \rangle \langle \Delta I(\lambda_2)^2 \rangle}}$, where $\langle \rangle_N$ denotes ensemble average over distinct $N$ single-shot spectra, and $\Delta I = I - \langle I \rangle_N$. The correlation varies over the range $-1 < \rho < 1$. The correlation matrix is symmetric across the positive diagonal and shows the relationship between intensity variations at different wavelengths from shot-to-shot.

**Low-pass Fourier filtering** Because the average spectral envelope of the supercontinuum source varies as a function of wavelength in the measurements spectral window of a single realization, this distortion will be seen a part of the object by the ghost imaging scheme. In order to remove the spectral envelope distortion, we apply low-pass filtering of the correlation function in the Fourier space, i.e. we multiply the Fourier transform of the correlation function by a narrowband notch function centered over the central peak of the Fourier transform. This effectively suppresses the slowly-varying spectral envelope as compared to the rapidly-varying gas absorption features.

**Acknowledgments:** C.A. acknowledges the support from TUT and SPIM graduate schools. J.M.D. acknowledges support from the French Investissements d'Avenir program, project ISITE-BFC (contract ANR-15-IDEX-0003). G.G. acknowledges the support from the Academy of Finland (grant 298463).

**Author contributions:** C. A and P. R. constructed the experimental setup and conducted the experiments. All authors performed the data analysis and contributed to writing the manuscript. Overall project supervision was provided by G.G.

**Competing financial interests:** The authors declare that they have no competing financial interests.

**References**

[1] Erkmen B.I., Shapiro J.H. Ghost imaging: from quantum to classical to computational. Advances in Optics and Photonics **2**, 405-450 (2010).

[2] Padgett M.J., Boyd R.W. An introduction to ghost imaging: quantum and classical. Philosophical Transactions of the Royal Society A: Mathematical, Physical and Engineering Sciences **375**, 20160233 (2017).

[3] Bennink R.S., Bentley S.J., Boyd R.W., Howell J.C. Quantum and classical coincidence imaging. Physical Review Letters **92**, 033601 (2004).

[4] Pittman T.B., Shih Y.H., Strekalov D.V., Sergienko A.V. Optical imaging by means of two photon quantum entanglement. Physical Review A **52**, R3429 (1995).

[5] Bennink R.S., Bentley S.J., Boyd R.W. "Two-photon" coincidence imaging with a classical source. Physical Review letters **89**, 113601 (2002).

[6] Ferri F., Magatti D., Gatti A., Bache M., Brambilla E. *et al*. High-resolution ghost image and ghost diffraction experiments with thermal light. Physical Review Letters **94**, 183602 (2005).

[7] Sun B., Edgar M.P., Bowman R., Vittert L.E., Welsh S.S. *et al*. 3d computational imaging with single-pixel detectors. Science **340**, 844-847 (2013).

[8] Zhang D.J., Li H.G., Zhao Q.L., Wang S., Wang H.B. *et al*. Wavelength-multiplexing ghost imaging. Physical Review A **92**, 013823 (2015).

[9] Devaux F., Moreau P.A., Denis S. & Lantz E. Computational temporal ghost imaging. Optica **3**, 698-701 (2016).

[10] Gatti A., Brambilla E., Bache M., Lugiato L.A. Correlated imaging, quantum and classical. Physical Review A **70**, 013802 (2004).

[11] Shirai T., Setälä T. & Friberg A.T. Temporal ghost imaging with classical non-stationary pulsed light. J. Opt. Soc. Am. B **27**, 2549-2555 (2010).

[12] Ryczkowski P., Barbier M., Friberg A.T., Dudley J.M. & Genty G. Ghost imaging in the time domain. Nature Photonics **10**, 167-170 (2016).

[13] Ryczkowski P., Barbier M., Friberg A.T., Dudley J.M., Genty G. *et al*. Magnified time-domain ghost imaging. APL Photonics **2**, 046102 (2017).

[14] Goda K. & Jalali B. Dispersive fourier transformation for fast continuous single-shot measurements. Nature Photonics **7**, 102-112 (2013).

[15] Janassek P., Blumenstein S. & Elsäßer W. Ghost spectroscopy with classical thermal light emitted by a superluminescent diode. Physical Review Applied **9**, 021001 (2018).

[16] Dudley J.M., Genty G. & Coen S. Supercontinuum generation in photonic crystal fiber. Reviews of Modern Physics **78**, 1135-1184 (2006).

[17] Solli D.R., Herink G., Jalali B. & Ropers C. Fluctuations and correlations in modulation instability. Nature Photonics **6**, 463-468 (2012).

[18] Wetzel B., Stefani A., Larger L., Lacourt P.A., Merolla J.M. *et al*. Real-time full bandwidth measurement of spectral noise in supercontinuum generation. Scientific Reports **2**, 882 (2012).

[19] Ryczkowski P., Närhi M., Billet C., Merolla J.M., Genty G. *et al*. Real-time full-field characterization of transient dissipative soliton dynamics in a mode-locked laser. Nature Photonics **12**, 221-227 (2016).

[20] Wei Y., Li B., Wei X., Yu Y., Wong K.K. Ultrafast spectral dynamics of dual-color-soliton intracavity collision in a mode-locked fiber laser. Applied Physics Letters **112**, 081104 (2018).

[21] Solli D.R., Chou J. & Jalali B. Amplified wavelength–time transformation for real-time spectroscopy. Nature Photonics **2**, 48-51 (2007).

[22] Chou J., Solli D.R. & Jalali B. Real-time spectroscopy with subgigahertz resolution using amplified dispersive fourier transformation. Applied Physics Letters **92**, 111102 (2008).

[23] DeVore P.T.S., Buckley B.W., Asghari M.H., Solli D.R. & Jalali B. Coherent time-stretch transform for near-field spectroscopy. IEEE Photonics Journal **6**, 1-7 (2014).

[24] Saltarelli F., Kumar V., Viola D., Crisafi F., Preda F. *et al*. Broadband stimulated Raman scattering spectroscopy by a photonic time stretcher. Optics Express **24**, 21264-21275 (2016).

[25] Dobner S. & Fallnich C. Dispersive fourier transformation femtosecond stimulated Raman scattering. Applied Physics B **122**, 278 (2016)

[26] Mahjoubfar A., Churkin D.V., Barland S., Broderick N., Turitsyn S.K. *et al*. Time stretch and its applications. Nature Photonics **11**, 341-351 (2017).

[27] Hollas J.M. Modern Spectroscopy (Wiley, Chichester, 2004).

[28] Ferri F., Magatti D., Lugiato L.A. & Gatti, A. Differential ghost imaging. Physical Review Letters **104**, 253603 (2010).

[29] Sun B., Welsh S.S., Edgar M.P., Shapiro J.H. & Padgett M.J. Normalized ghost imaging. Optics Express **20**, 16892-16901 (2012).